# Optimized Quality Factor of Fractional Order Analog Filters with Band-Pass and Band-Stop Characteristics

Anindya Pakhira[1], Saptarshi Das[2], Anish Acharya[1]

1. Department of Instrumentation and Electronics Engineering, Jadavpur University, Salt-Lake Campus, LB-8, Sector 3, Kolkata-700098, India.
2. Department of Power Engineering, Jadavpur University, Salt-Lake Campus, LB-8, Sector 3, Kolkata-700098, India.
Email: saptarshi@pe.jusl.ac.in

Indranil Pan[2,3], Suman Saha[4]

3. MERG, Energy, Environment, Modelling and Minerals ($E^2M^2$) Research Section, Department of Earth Science and Engineering, Imperial College London, Exhibition Road, London SW7 2AZ, UK.
4. Drives and Control System Technology Group, CSIR-Central Mechanical Engineering Research Institute, Mahatma Gandhi Avenue, Durgapur-713209, India.

*Abstract*—**Fractional order (FO) filters have been investigated in this paper, with band-pass (BP) and band-stop (BS) characteristics, which can not be achieved with conventional integer order filters with orders lesser then two. The quality factors for symmetric and asymmetric magnitude response have been optimized using real coded Genetic Algorithm (GA) for a user specified center frequency. Parametric influence of the FO filters on the magnitude response is also illustrated with credible numerical simulations.**

## I. Introduction

Band pass and band-stop filters are perhaps the most widely used family of analog filters, with applications in a wide variety of domains. Besides their heavy use in communication, BP filters are used in audio circuits, and also have applications in instrumentation, seismology, sonar, optics and medical equipment technologies. BS filters are used for signal rejection in communications, audio, optics, etc. The modern applications of BP and BS filters require very precise filtering, in order to remove interference from other closely spaced frequency bands. In other words the quality factors (Q-Factor) of the filters need to be considerably high [1]-[2]. Recently, with the successful realization of non-integer order fractance devices, there has been a growing interest in employing fractional order filters in the performance-intensive applications [3]. The fractional order filters give additional control over the filter characteristics as compared to their integer order counterparts.

Electronic filters have traditionally been implemented using capacitors and inductors [1]-[2]. However, FO filters can be considered to be a special combination of a FO electrical element known as fractance, the transfer function of which is given by (1) [4]-[7]. In the range, $0 < \alpha < 2$ it represents what is known as an FO inductor, while in the range $-2 < \alpha < 0$ it represents an FO capacitor. The standard inductor and capacitor are represented in the cases $\alpha = 1, \alpha = -1$ respectively. For $\alpha = -2$, it represents the frequency dependent negative resistance (FDNR). Thus fractances are the basic FO elements which can be used to realize fractional order filters in real electronic circuits [8]-[10].

$$Z(s) = as^{\alpha} \Rightarrow Z(j\omega) = a\omega^{\alpha} e^{j(\pi\alpha/2)}, \quad \alpha \in [-2, 2] \quad (1)$$

In the integer-order paradigm, BP and BS filters require at least a second order transfer function [1]. However, BP filters in fractional order domain, due to the added flexibility of manipulating the slopes corresponding to each pole or zero, can be implemented in an order even less than two [4]. In the present work, an FO BP filter with order less than two has been optimized to obtain band-pass and band-stop characteristics for specified center frequency. The Q-factors have been optimized here with real coded Genetic Algorithm to obtain enhanced filtering action. Simulation studies have been carried out for designing optimum fractional order BP and BS filters with symmetric and asymmetric magnitude responses respectively. Fractional order filters are recently being popular in recent literatures like fractional Butterworth filter [11], for high quality factor [12], fractional low pass filtering [13], in biomedical applications [14], fractional lead filter design [15], approximation of fractional Laplacian operator [16], fractional second order filter approximation [17], fractional step-filter [18] etc., which motivates us for the investigation of optimized quality factor of fractional order filters having order less than two, producing symmetric/asymmetric magnitude curve.

The rest of the paper is organized as follows. Section II describes the FO band-pass filter, its parametric influence and optimum Q-factor design. Section III extends the proposed concepts for the design of FO band-stop filters, also with an optimization framework. The paper ends with the conclusion and future directions in section V, followed by the references.

## II. FO Band-pass Filters with Order Less Than Two

Radwan *et al.* [4] first proposed the generalized transfer function of a FO band pass filter as (2) with the pole and zero parameters being $\{a, b\}$ and associated fractional orders being $\{\alpha, \beta\}$. Expression for the magnitude response of the filter is given by (3). It is clear that the filter gives band-pass characteristics for $\alpha > \beta$. Also the magnitude curve of the BP filter (2) is symmetric for $\alpha = 2\beta$ [4]. In the pioneering literature [4], simulation studies have been shown for pole-zero parameters $a=1, b=1$. This paper maximizes the Q-factor of the BP filter with real-coded GA based optimization approach to obtain optimum filter parameters i.e. $\{a,b,\alpha,\beta\}$.

$$T_{FBPF}(s) = \frac{bs^\beta}{s^\alpha + a} \Rightarrow T_{FBPF}(j\omega) = \frac{b(j\omega)^\beta}{(j\omega)^\alpha + a} \qquad (2)$$

$$|T_{FBPF}(j\omega)| = \left| \frac{b\omega^\beta \left( \cos\left(\frac{\beta\pi}{2}\right) + j\sin\left(\frac{\beta\pi}{2}\right) \right)}{\omega^\alpha \left( \cos\left(\frac{\alpha\pi}{2}\right) + j\sin\left(\frac{\alpha\pi}{2}\right) \right) + a} \right| \qquad (3)$$

$$= \frac{b\omega^\beta}{\sqrt{\omega^{2\alpha} + 2a\omega^\alpha \cos\left(\frac{\alpha\pi}{2}\right) + a^2}}$$

Radwan *et al.* [4] have given detailed analysis to show that FO filter of the form (2) exhibits symmetric band-pass characteristics if the denominator order be twice of the numerator order i.e. $\alpha = 2\beta$. Therefore, the frequency corresponding to the maximum value of the magnitude response where its peak occurs ($\omega_m$) is equal to the center frequency ($\omega_0$). The parametric influence of the symmetric FO band-pass filters has been shown in Fig, 1-3 with respect to the nominal pole/zero parameters adopted from [4] as $a=1, b=1$. In all frequency domain simulation studies of FO filters, presented in this paper, the FOTF Toolbox introduced in [19] has been used. From the magnitude responses with varying fractional order ($\beta$) of symmetric band-pass filters in Fig. 1, it is evident that the gain roll-off is high for higher values of the filter order $\beta$, but the maximum peak occurs near $\beta = 1$.

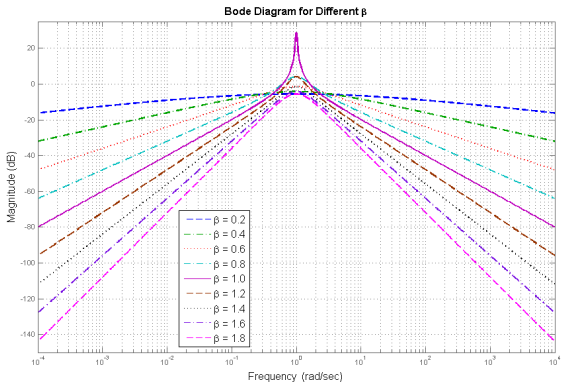

Figure 1. Effect of numerator order (β) on the symmetric BP filter response.

Fig. 2 shows that higher value of the pole parameter ($a$) shifts the peak of the BP filter towards higher frequencies and also the magnitude at $\omega_m$ reduces gradually indicating reduced band-pass performance. Also, higher value of zero parameter ($b$) pushes the magnitude upward i.e. towards high value (quality factor) at the same $\omega_m$, while shift of the peak/centre frequency does not occur. Thus it is logical to conclude From Fig. 1-3 that low value of the pole parameter ($a$) and high value of the zero parameter ($b$) with fractional order somewhere near unity is capable of producing high quality of band-pass performance for specified centre frequency ($\omega_0$).

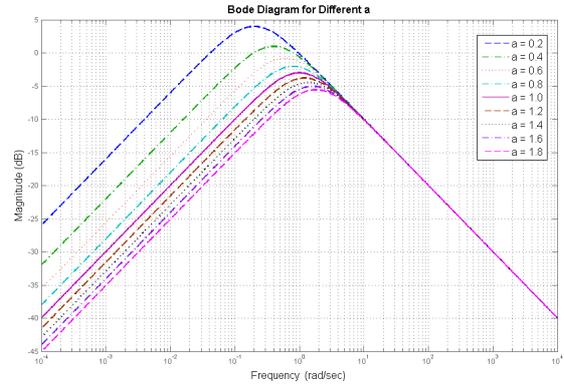

Figure 2. Effect of pole parameter (a) on the symmetric BP filter response.

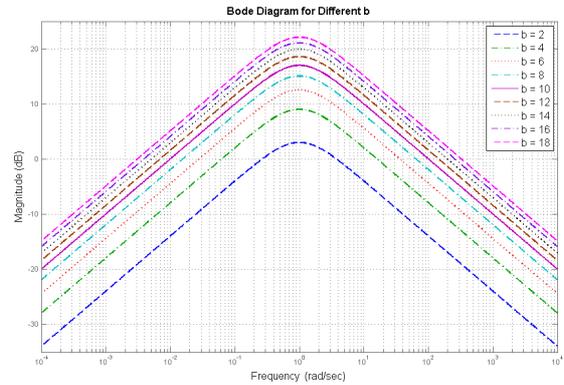

Figure 3. Effect of zero parameter (b) on the symmetric BP filter response.

In classical theory of analog filter design, there is no concept of quality factor for first order filters. Conventional theory shows that the quality factor ($Q$) for a specified center frequency ($\omega_0$) can only be found for biquad filters [1]. Here, in order to design symmetric/asymmetric FO band-pass filters with an optimization based technique another concept of Q-factor has been used. It is reported in [20] that the gain of a filter at exactly its center frequency is equal to its quality factor ($Q$). In symmetric FO filters, due to the fact $\omega_m = \omega_0$, its quality factor can then be represented as:

$$T_{FBPF}(j\omega)\big|_{\omega=\omega_0} = \frac{b\omega_0^\beta}{\sqrt{\left(\omega_0^{2\alpha} + 2a\omega_0^\alpha \cos\frac{\alpha\pi}{2} + a^2\right)}} = Q \qquad (4)$$

From (4), it is evident that the Q-factor is a function of the filter parameters $\{a,b,\alpha,\beta\}$ for a user specified peak (center)

frequency $\omega_0$. To maintain the symmetry of the magnitude response of the filter, the independent parameters influencing the quality factor are chosen as $\{a, b, \beta\}$ only.

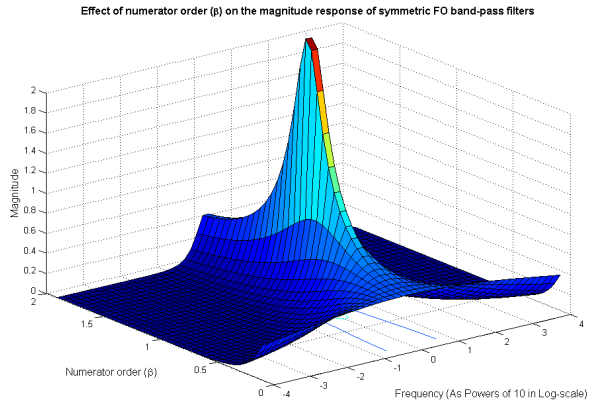

Figure 4. Effect of numerator order (β) on the magnitude of symmetric BPF.

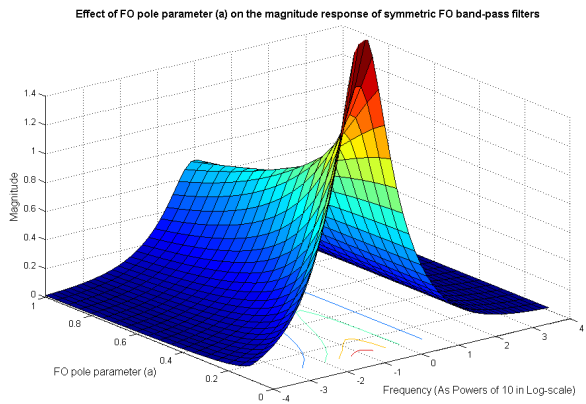

Figure 5. Effect of pole parameter (a) on the magnitude of symmetric BPF.

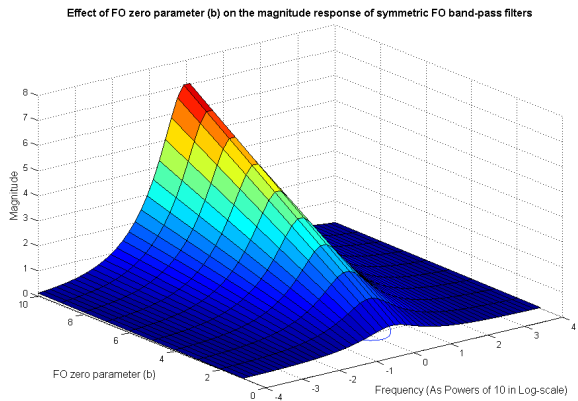

Figure 6. Effect of zero parameter (b) on the magnitude of symmetric BPF.

Fig. 4 shows the variation in the magnitude frequency response for symmetric BP filters ( $\alpha = 2\beta$ ) with the numerator order $\beta$, considering other parameters as constants i.e. $a = b = 1$, as studied in Radwan *et al.* [4]. Next, variation of the magnitude frequency responses of the FO band-pass filter have been shown with respect to the pole parameter and zero parameter in Fig. 5 and Fig. 6 respectively. It is clear that the magnitude increases for low value of "$a$" and high value of "$b$" and fractional order near unity as also found from Fig. 1-3. Therefore, the optimization space will have combined influence of the decision variables $\{a, b, \beta\}$, as depicted in Fig. 4-6. From the 3-dimensional magnitude frequency responses with respect to the filter parameters, it is evident that though the individual variation in $Q$ is smooth, the combined landscape of the search space may be quite complicated and may have local minima. This motivates the application of population based intelligent optimization algorithm like GA over the classical gradient based methods in order to ensure that the true global minima has been found in the optimization process. In the present Q-factor maximization problem, real-coded GA is used. A solution vector of the decision variables $\{a, b, \beta\}$ is initially randomly chosen from the search space and undergoes reproduction, crossover and mutation, in each iteration to give rise to a better population of solution vectors in further iterations. Here, the number of population members in GA is chosen to be 20. The crossover and mutation fraction are chosen to be 0.8 and 0.2 respectively [21] for maximization of the quality factor. Real-coded GA based optimization has been attempted next in order to maximize the Q-factor of the FO band-pass filter given by (4) within the parametric bounds $\beta \in [0, 2]$ and $\{a, b\} \in [0, 20]$. The algorithm has been run several times for a specified center frequency $\omega_0 = 1.5 \ rad/\sec$ and the best results are reported in (5) along with the corresponding magnitude curve shown in Fig. 4.

$$Q = 22.6017, a = 0.996307, b = 18.2033, \beta = 0.924351 \quad (5)$$

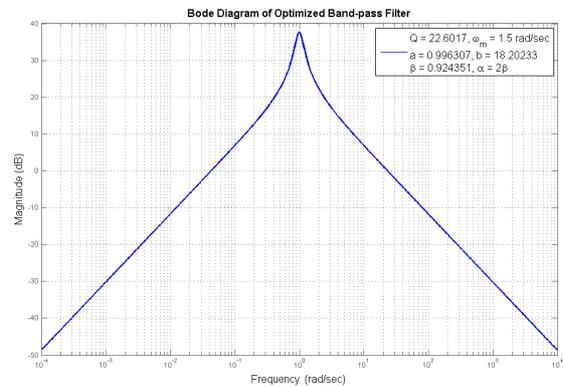

Figure 7. Magnitude response of optimized symmetric FO band-pass filter.

III. FO BAND-STOP FILTERS WITH ORDER LESS THAN TWO

Similar treatment gives the generalized transfer function of a FO band-stop filter as (6) with the pole and zero parameters being $\{b, a\}$ and associated fractional orders being $\{\beta, \alpha\}$. It is evident that the band-stop filter is just an inverse of the FO band-pass filter transfer function (2). Expression for the magnitude response of the filter is given by (7) and for the Q-factor in (8). This typical filter structure gives band-stop characteristics for $\alpha > \beta$. Also the magnitude curve of the FBSF (7) is symmetric for $\alpha = 2\beta$, as also in the FBPF cases. The parametric influences of the fractional band-stop filter have been shown in Fig. 8-10, as three dimensional magnitude frequency responses in dB scale.

$$T_{FBSF}(s) = \frac{s^{\alpha} + a}{bs^{\beta}} \Rightarrow T_{FBSF}(j\omega) = \frac{(j\omega)^{\alpha} + a}{b(j\omega)^{\beta}} \quad (6)$$

$$\left| T_{FBSF}(j\omega) \right| = \sqrt{\omega^{2\alpha} + 2a\omega^{\alpha} \cos\left(\frac{\alpha\pi}{2}\right) + a^2} \Big/ \left( b\omega^{\beta} \right) \quad (7)$$

$$\left. T_{FBSF}(j\omega) \right|_{\omega = \omega_0} = \frac{\sqrt{\left( a^2 + \omega_0^{2\alpha} + 2a\omega_0^{\alpha} \cos\frac{\alpha\pi}{2} \right)}}{b\omega_0^{\beta}} = \frac{1}{Q} \quad (8)$$

To ensure that the FO filter gives optimum band-rejection performance GA has also been employed to maximize the Q-factor (8) of the FO band-stop filter given by (6). The best found optimization result for a specified center frequency of $\omega_0 = 1.5 \; rad/\sec$ has been reported in (9), along with the corresponding magnitude curve shown in Fig. 11.

$$Q = 21.2739, a = 0.99767, b = 17.11228, \beta = 0.92593 \quad (9)$$

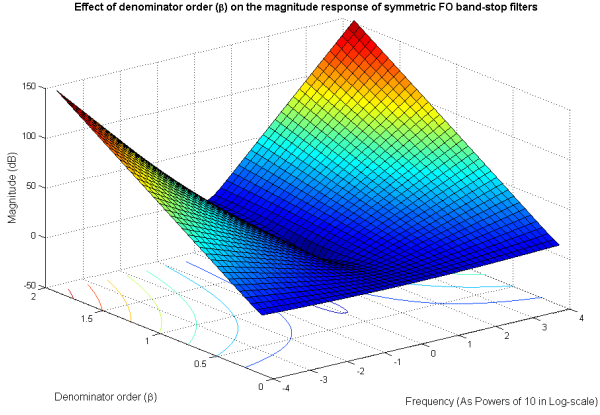

Figure 8. Effect of denominator order (β) on magnitude of symmetric BSF.

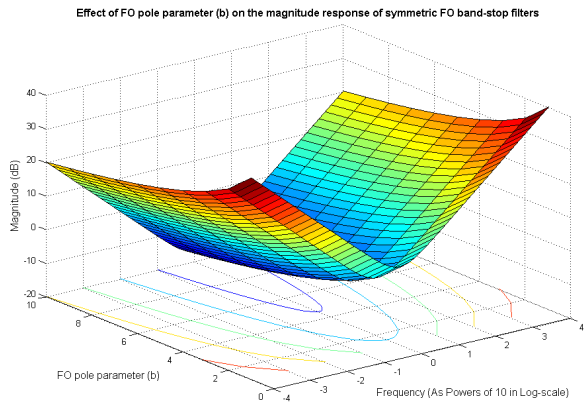

Figure 9. Effect of pole parameter (b) on the magnitude of symmetric BSF.

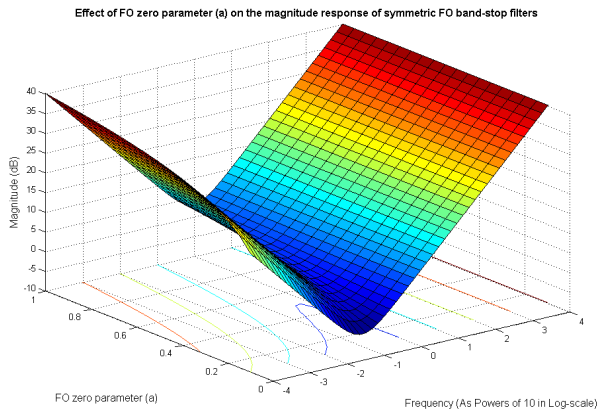

Figure 10. Effect of zero parameter (a) on the magnitude of symmetric BSF.

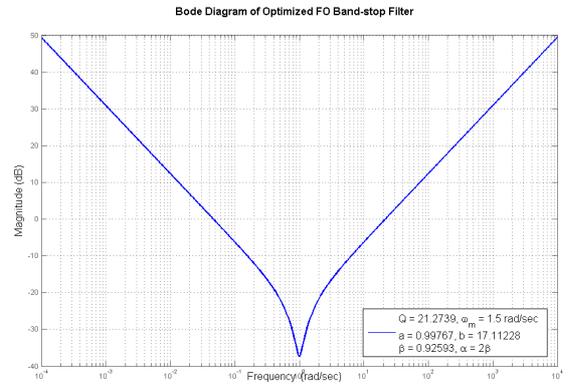

Figure 11. Magnitude response of optimzed symmetric FO band-stop filter.

## IV. QUALITY FACTOR OPTIMIZATION OF FRACTIONAL BP AND BS FILTERS WITH ASYMMETRIC MAGNITUDE RESPONSE

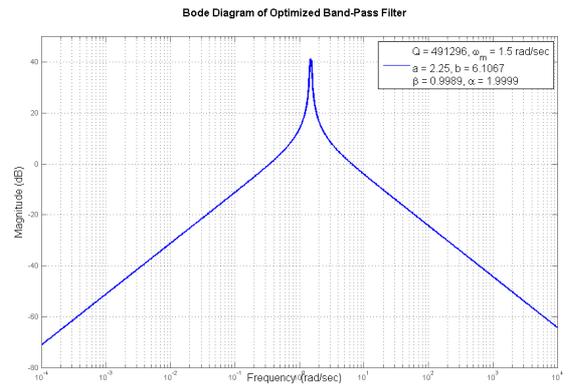

Figure 12. Magnitude response of optimized asymmetric FO band-pass filter.

It is indeed logical that the fractional filters may produce high quality factors with additional flexibility of choosing its numerator and denominator orders separately, thus producing an asymmetric magnitude curve. The proposed optimization based Q-factor tuning approach has been applied to design fractional filters with band-pass and band-stop characteristics. It is found that asymmetric filters are capable of producing very high quality factor in comparison with the symmetric fractional filters as shown Fig. 12-13, with the respective optimum filter parameters and optimized Q-factor.

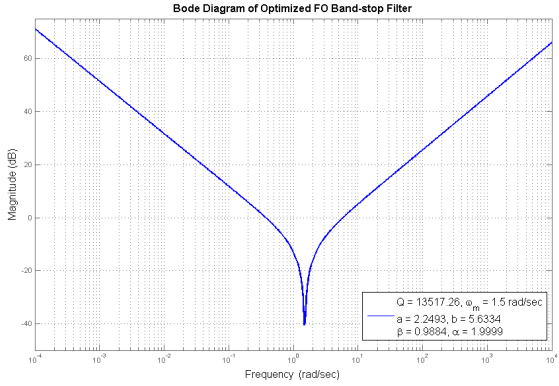

Figure 13. Magnitude response of optimized asymmetric FO band-stop filter.

Similar population based optimization approach of fractional filter design can be found in recent literatures for rational approximation [22]-[24] and filtering applications [25].

## V. QUALITY FACTOR OPTIMIZATION FOR FRACTIONAL SECOND ORDER FILTERS

Radwan *et al.* [5] has also given the generalized fractional second order filter formulation, similar to the fractional first order filer [4], which has been optimized for enhanced quality in the previous sections. The transfer function and magnitude frequency response of the fractional second order band-pass filter is given by (10) and (11) respectively. It is evident from the filter structure itself, that upon optimization the second term of the denominator vanishes to produce a spike in the frequency domain. To verify the argument, the GA based quality optimization has been applied for the fractional second order structure (10) also. The optimization with high quality factor using (11) enforces the parameter "$a$" in the denominator of (10) to tend towards zero, thus diminishing the fractional second order filter structure to produce the symmetric fractional first order filter. Due to the above mentioned shortcomings of fractional second order structure (10) as proposed in [6], its parametric optimization for high quality is not recommended.

$$T_{FBPF-II}(s) = \frac{ds^\alpha}{s^{2\alpha} + 2as^\alpha + b}$$

$$\Rightarrow T_{FBPF-II}(j\omega) = \frac{d(j\omega)^\alpha}{(j\omega)^{2\alpha} + 2a(j\omega)^\alpha + b} \quad (10)$$

$$|T_{FBPF-II}(j\omega)| = \left| \frac{d\omega^\alpha \left( \cos\left(\frac{\alpha\pi}{2}\right) + j\sin\left(\frac{\alpha\pi}{2}\right) \right)}{\omega^{2\alpha}\left(\cos(\alpha\pi) + j\sin(\alpha\pi)\right) + 2a\omega^\alpha \left(\cos\left(\frac{\alpha\pi}{2}\right) + j\sin\left(\frac{\alpha\pi}{2}\right)\right) + b} \right|$$

$$= \frac{d\omega^\alpha}{\sqrt{\omega^{4\alpha} + 4a\omega^{3\alpha}\cos\left(\frac{\alpha\pi}{2}\right) + \left(4a^2 + 2b\cos(\alpha\pi)\right)\omega^{2\alpha} + 4ab\omega^\alpha \cos\left(\frac{\alpha\pi}{2}\right) + b^2}} \quad (11)$$

## VI. CONCLUSION

Real-coded GA based quality factor tuning has been studied in this paper for FO generalization of first order filters to obtain optimum band-pass and band-stop performance even for orders less than two. The parametric influences of FO filters have been illustrated to give a better insight in the FO filter design. It has also been found that the second order generalization of fractional band-pass filters upon Q-factor optimization gives impractical results due to the fact that in order to produce increasingly high Q-factor, the first pole parameter ($a$) as in (10) approaches towards zero. Therefore, the fractional second order filter structure in [5] reduces to the fractional generalization of the first order filters (2). More work is needed in this area, and comparison with equivalent optimum digital filtering techniques also needs to be pursued. In future, optimization of irrational filters, having fractional power of a rational transfer function like [26], [17] can be investigated.